# What Human-Horse Interactions may Teach us About Effective Human-AI Interactions

Mohammad Hossein Jarrahi* and Stanley Ahalt

University of North Carolina at Chapel Hill

This article explores human-horse interactions as a metaphor for understanding and designing effective human-AI partnerships. Drawing on the long history of human collaboration with horses, we propose that AI, like horses, should complement rather than replace human capabilities. We move beyond traditional benchmarks such as the Turing test, which emphasize AI's ability to mimic human intelligence, and instead advocate for a symbiotic relationship where distinct intelligence enhances each other. We analyze key elements of human-horse relationships—trust, communication, and mutual adaptability—to highlight essential principles for human-AI collaboration. Trust is critical in both partnerships, built through predictability and shared understanding, while communication and feedback loops foster mutual adaptability. We further discuss the importance of taming and habituation in shaping these interactions, likening it to how humans train AI to perform reliably and ethically in real-world settings. The article also addresses the asymmetry of responsibility, where humans ultimately bear the greater burden of oversight and ethical judgment. Finally, we emphasize that long-term commitment and continuous learning are vital in both human-horse and human-AI relationships, as ongoing interaction refines the partnership and increases mutual adaptability. By drawing on these insights from human-horse interactions, we offer a vision for designing AI systems that are trustworthy, adaptable, and capable of fostering symbiotic human-AI partnerships.

## 1 INTRODUCTION

AI systems are progressively making inroads into more application contexts. However, there is a growing consensus that replacing humans who have specialized expertise and implicit knowledge with AI systems may not be necessarily beneficial [26]. Some conventional benchmarks for AI, such as the Turing test, might not fully capture the roles AI can play alongside human intelligence since they encourage AI to mimic and implicitly replace human intelligence rather than complement it [13]. Instead of full human emulation by AI, humans should remain "in the loop" in many processes, suggesting that the future of computing depends on a collaborative partnership between human and AI systems, rather than a competition between them [28]. But the question remains: how can this partnership be envisioned, considering the different types of intelligence that humans and AI possess?

Understanding and envisioning this partnership often requires using metaphors that are familiar and inspiring to us. This approach is an example of analogical reasoning [10], where we understand a less familiar or emerging concept (such as human-AI partnership) by comparing it to a more familiar idea (like the human-horse partnership). Metaphors also play a pivotal role in envisioning how we design and interact with AI systems [20]. In this context, we suggest that the millennia of successful interaction and partnership that humans cultivated with horses, which transformed transportation, food production, and warfare, can provide a familiar metaphor and practical insights, and it makes sense to consider AI and horses analogously.

* Corresponding author: jarrahi@unc.edu

This approach offers a framework for designing symbiotic interactions between humans and AI by drawing on long-standing principles of complementary intelligence found in human-animal interactions, particularly focusing on how these principles can inform AI design. First, horses embody a different form of intelligence, one that has complemented and augmented humans' capabilities for millennia [19]. By drawing our well-established mental model in interacting with horses, we can gain valuable insights into how to collaborate effectively with AI, with the goal of augmenting our own intelligence. Second, by analogizing horses and our relationship with them, we can possibly produce a more accurate representation of how AI may complement humans [23], indicating fully emulating human behavior in AI systems, as implied in benchmarks like the Turing test, may not always be the most synergistic approach when designing and implementing AI systems.

In this article, we will outline various ways in which human-horse interactions can serve as a model for imagining methods to create, sustain, and amplify human-AI partnerships.

## 2 AI AND HORSES AS PARTNERS, NOT REPLACEMENTS FOR HUMANS

Animal analogies have been already used to describe our interactions with AI or robots, presenting intelligent machines as partners, not competitors, and highlighting their role in complementing, not replacing, human contributions [e.g., 8,23]. These analogies also help us imagine and employ AI as a complement to human labor, rather than tools that mimic and automate humans away. The analogy of human-horse interaction offers a parallel for understanding 'human-AI symbiosis' [15]. In unison, humans and horses can perform music in dressage events, or under a rider's direction, horses can achieve jumps of 27 feet in length. The performance of wild horses, such as leaping across a stream, pales in comparison to the athletic achievements of trained horses in competitive jumping events, where they navigate multiple obstacles over short distances at high speeds. Therefore, as Janet Jones, a neuroscientist and horse trainer, points out, "comparing the wilderness hop with the show jump is like associating a flintstone with a nuclear bomb" [17].

As prey animals, horses have pupils that allow for clear vision along the horizon but offer blurred vision above and below it. Furthermore, horses, being fear-flight animals, have limited capacity for interpretation, unlike humans, who are capable of devising, executing, and revising plans [31]. In contrast, human anatomy and brain functions have evolved for predation, providing us with depth perception, sharp focus, and an advanced prefrontal cortex. These features enable strategic thinking and reasoning [17]. Similarly, humans and AI can be seen as natural allies partly because of our synergistic differences: these contrasting qualities complement each other, creating a synergy that makes the combination of humans and AI more potent than alone. For instance, AI excels in data analysis, processing vast amounts of sensory data, while humans offer strategic insight and a holistic view [3]. This enables them to contextualize sensory insights and communicate and drive decisions with multiple stakeholders.

## 3 TRUST AS A KEY INGREDIENT BETWEEN THE PARTNERS

Trust is a critical factor in the success of human-AI interactions, influencing user acceptance and the integration of AI contributions into human decision-making processes. Studies highlight the ways in which humans can develop trust in their animal partners, such as being able to predict the animal's actions in various situations [23]. Specifically trust arises only once there is a developed means of communication and sharing between the two parties [2]. Mutual trust means that humans often rely on the judgment and skills of horses, while horses rely on the guidance and care of humans [19].

Effective riding requires riders to comprehend why their horses behave as they do. Riders should be able to develop a frame of mind of their horses' internal motivations or external factors influencing their behavior [14]. This understanding enables riders to respond more adaptively, debug issues with their horses when things go wrong, and retrain them in future interactions.

Similarly, effective interaction with AI necessitates a transparent AI design, ensuring users to some extent comprehend the system's decision-making process to foster trust. Furthermore, AI systems should exhibit consistent and predictable behavior in similar situations, enhancing user trust by reflecting the trust dynamics seen in human-animal relationships. However, this stands in contrast to the nature of many general-purpose generative AI systems, such as ChatGPT, which are intelligent yet still opaque and can deliver inconsistent performance [6]. This inconsistency potentially instills an element of surprise and distrust particularly in high-stakes decision-making scenarios. A partial solution here could be dedicating more attention to focused models trained on high-quality, domain-specific data, which could perform more effectively and reliably in specialized tasks [16]. The use of domain-specific models could result in practical success in various specialized contexts and possibly lead to greater user trust in AI performance.

## 4 TAMING AND HABITUATION OF BOTH HORSES AND AI

Interacting with AI closely and effectively involves a dual process of taming and habituating. In this analogy, while taming concentrates on reducing negative impacts, harm, or disorderly behavior resulting from the power of AI or horses [9], habituation focuses on strategically channeling that power in designing new workflows. In horse training, the initial taming phase is crucial as it involves teaching a wild or untrained "animal" to accept and trust human presence and guidance, acclimatizing it to human interaction. In a parallel manner, the initial training of AI involves a similar 'taming' phase, which could involve managing AI and its adverse or unintended effects, addressing the ethical issues it may raise prior to successfully collaborating with and leveraging these systems. This approach may require organizations to espouse a more careful and considered adoption strategy, one that explicitly focuses on ethical concerns related to biased algorithms that might threaten the privacy of different stakeholders, disrupt the balance between work and life, and undermine trust in the workplace.

Just as horses need to undergo a carefully orchestrated process of "habituation" to the human environment as part of their training [19], AI, in adapting to living "alongside" humans, must be fine-tuned to work within the complexities of the operational environment. AI systems need to be made adaptable to the realities outside controlled, laboratory settings. Many AI systems perform well in controlled, experimental environments but struggle in less predictable, real-world scenarios with multiple stakeholders. Consequently, AI systems must be customized and trained to function effectively across various human environments and contexts, such as health care, finance, education, and entertainment, ensuring its applicability in everyday life. This may include adapting to new situations, organizational processes, and the specific requirements set by the operating environment.

Recent developments, such as adversarial training and transfer learning, offer new ways to 'tame' and adapt AI systems to unpredictable real-world scenarios. Adversarial training, which involves training AI models to handle deceptive or misleading data inputs, enhances robustness and safety, making the systems more reliable in dynamic environments [11]. Transfer learning enables AI to apply knowledge learned from one domain to another [22], similar to how horses might transfer training from one discipline, like dressage, to another, like show jumping, showcasing adaptability and versatility.

## 5 MUTUAL ADAPTABILITY WITH CONTINUOUS COMMUNICATION

In an effective partnership between horses and humans, the key lies in understanding, patience, and adaptation. Horses adapt to their riders, and riders adapt to their horses. This mutual adaptation requires observation, sensitivity to feedback, and adjustments in behavior [12]. Ideally, AI systems should present the same level of adaptability, learning from each interaction to enhance future responses. This involves machine learning algorithms that adjust based on user behavior and preferences. On the human side, users adapt to AI by understanding its capabilities, tendencies and limitations, learning how to interact with it effectively.

Effective continuous communication is essential for adaptability in training and working with horses, operating as a two-way street: a rider's command (e.g., through hand and leg movements, seat positioning, or vocalization) provide feedback for the horse, while the horse's reactions serve as feedback for the rider [4]. This loop allows for mutual understanding and improvement while horses stay very responsive to their rider (for example horse's ears can communicate signals such as attentiveness to the rider). In interacting with AI, a similar feedback loop is essential. User responses and interactions with AI provide data that the AI system can use to learn and improve. At the same time, AI systems should provide feedback to users, helping them understand how to better interact with the system.

Communication goes beyond just initial training, mirroring the real-time interactions between humans and horses, where both are attuned to subtle environmental cues. Horses have the ability to perceive and respond to the slightest visual and auditory signals from their surroundings [30]. Likewise, AI systems are adept at monitoring customer chats and providing real-time response suggestions to human agents, effectively acting on the information presented to them [5].

## 6 SHARED AND HARMONIOUS DECISION MAKING

In horse-human dynamics, while humans generally guide the interaction, horses often have a degree of autonomy and can make decisions, especially in situations where they might sense danger or discomfort. This collaborative approach can lead to a more harmonious and effective partnership. In AI, this translates to systems designed to not only follow commands but also to provide suggestions, insights, or warnings based on their processing of data and sensing of the operating environment. This means AI should not be just seen as a passive tool but an active participant in the decision-making process, offering alternatives or cautioning against certain choices based on its own analytical frame.

To foster growth and enhance decision-making, both horses and AI systems require a degree of autonomy. Just as horses learn more effectively by having control over their actions and outcomes [7], AI systems similarly benefit from the liberty to make decisions and learn from the consequences of these decisions. This approach not only facilitates learning but also promotes a sense of partnership and shared decision making. Wipper [31] observes that horses demonstrate remarkable abilities when allowed to approach challenges in their own way. Like horses, AI systems should be afforded opportunities for trial and error. This method encourages a shared and harmonious decision-making environment where both humans and AI can expand their learning and contribute meaningfully to the workflow.

One practical example of such a collaboration between AI and humans can be found in medical diagnostics, in which AI systems have been implemented to assist radiologists by highlighting areas of concern on medical images. These systems do not replace radiologists but rather enhance their ability to diagnose more accurately and quickly [29]. Similarly, in the context of autonomous vehicles, AI systems work in tandem with human drivers, providing real-time feedback and taking control in specific scenarios, such as highway driving or emergency braking, to ensure safety and improve decision-making outcomes [27]

## 7 ASYMMETRIC RESPONSIBILITY AND CONTINUOUS HUMAN SUPERVISION

As noted, riders and their horses frequently share a bond of mutual trust. This trust, however, does not negate the need for awareness of the horse's potential for unpredictable behavior, driven by instinctual reactions such as shying away or galloping off. The unpredictable behaviors of horses mirror the challenges faced by users of AI systems, which due to their capacity for self-learning, can also produce unforeseen and unpredictable outcomes. In both scenarios, the key lies in readiness to vigilantly look for and respond to these unpredictable behaviors.

The establishment of clear boundaries and guidelines is crucial in managing these relationships effectively. In the context of horse training, tools like the "rein" symbolize a method of communication that reinforces human authority and responsibility. Similarly, in the realm of AI, maintaining human oversight, or "staying in the loop," through clear guidelines and regulations—often referred to as AI guardrails—ensures the safety, ethics, and beneficial use of AI technologies. This interaction hierarchy, where humans generally hold authority over animals and AI systems, suggests the importance of mitigating potential risks through clear boundaries and oversight [21]. For example, in autonomous drones, the AI system may take initial assessments and navigation, but human operators are always in the loop, making strategic decisions based on the AI's feedback [1]. This reflects the balance between autonomy and supervision, where AI handles heavy data tasks, and humans provide oversight and context-sensitive judgment, ensuring both efficiency and ethical responsibility.

Acknowledging the inherent asymmetry of responsibility in both relationships is vital to a human-centered approach [28]. That is, despite the collaboration and mutual learning in these partnerships, humans bear the ultimate responsibility and authority, steering the general direction and ensuring that both horses and AI systems align with their broader objectives and safety protocols. In horse-human partnerships, humans must assume the leadership role, guiding the horse while allowing it the liberty to make decisions within set boundaries. It is a delicate balance of granting autonomy and maintaining control, embodying the paradox that, although the horse may have the freedom to think for itself, it ultimately must align with the human's directives and objectives [31]. Similarly, interactions with AI systems, lower on consciousness or rational autonomy [24], present a fundamentally asymmetric relationship, which places the burden of responsibility on humans, including those not directly interfacing with the AI.

## 8 LONG-TERM COMMITMENT AND LEARNING

After the initial phase of training horses, the focus naturally transitions to the ongoing maintenance and the potential for enhancing their training. This evolution underlines the necessity of a long-term commitment to successful horse-human relationships, which are sustained through continuous care and training. In a similar manner, the development and integration of AI mirror this commitment, necessitating regular maintenance, timely updates, and stringent ethical oversight. AI systems, fundamentally reliant on data-driven training, require continuous learning to improve their performance and responsiveness. These dynamic mandates human oversight, ensuring diligent management and supervision to maintain control.

In addition, the symbiosis in a horse-human relationship is a testament to mutual learning and growth. In these partnerships, both the horse and the human learn from each other, fostering a bond that leads to significant improvement in their interactions [18]. This concept of mutual learning extends to human-AI interactions, advocating for a similar balance. Here, the AI system enhances its performance by learning from the user's behavior, while the user, in turn, becomes more adept at interacting with the AI, paving the way for a mutually beneficial relationship beyond simply automating tasks.

The equine brain can be described as a formidable learning machine capable of understanding various human cues, which significantly influences the horse-rider interaction [17]. However, this learning process requires strategic patience, emphasizing the importance of carefully monitoring the horse's readiness for new tasks. Prematurely overmatching a horse can detrimentally affect their potential for future successes [31]. This principle of measured progression and mutual learning applies not only to equine training but also to AI integration. In both contexts, a focus on continuous improvement, ethical management, and fostering synergistic relationships between humans and technology is crucial for long-term commitment and success. This approach may contrast with current uses of AI, which are often primarily directed at achieving short-term gains and efficiency goals through automation.

## 9 CONCLUDING REMARKS

In conclusion, the integration of human and AI’s unique capabilities heralds a new era of collaboration, much like the age-old partnership between humans and horses. This analogy provides a lens that can help us consider human-AI interactions from an informed, ethical, and potentially perspective, potentially setting a foundation for building AI systems that aim to be trustworthy, adaptable, and harmonious with human values and objectives. Therefore, our goal should not be to engineer machines that mimic human cognition but to develop AI systems that augment and complement human intelligence, fostering a partnership where the strengths of one compensate for the weaknesses of the other. Such a synergistic relationship could evolve into a centaur-like (half-human, half-horse mythological creatures) dynamic [25], blurring the lines between human and AI, and highlighting the superiority of a united human-AI team over two entities working alone [19]. Inspired by seamless interaction between rider and horse, we envision a future where the hybridization of human and AI can achieve unparalleled coherence, interdependence, and complementarity, effectively creating a powerful ‘centaur’ force for organizational and business success.